\documentclass[twocolumn,groupedaddress,superscriptaddress]{revtex4}
\usepackage{eurosym}
\usepackage{graphicx}
\usepackage{amssymb}
\usepackage{latexsym}
\usepackage[usenames]{color}
\usepackage{float}
\usepackage{hyperref}

\begin{document}

\title{Quantum avalanche in the Fe$_{8}$ Molecular-Magnet}
\author{Tom Leviant}
\affiliation{Department of Physics, Technion - Israel Institute of Technology, Haifa,
32000, Israel}
\author{Eli Zeldov}
\affiliation{Department of Condensed Matter Physics, The Weizmann Institute of Science,
76100 Rehovot, Israel}
\author{Yuri Myasoedov }
\affiliation{Department of Condensed Matter Physics, The Weizmann Institute of Science,
76100 Rehovot, Israel}
\author{Amit Keren}
\affiliation{Department of Physics, Technion - Israel Institute of Technology, Haifa,
32000, Israel}
\date{\today }

\begin{abstract}
We report spatially resolved, time-dependent, magnetization reversal
measurements of an Fe$_{8}$ single molecular magnet using a microscopic Hall
bar array. We found that under some conditions the molecules reverse their
spin direction at a resonance field in the form of an avalanche. The avalanche
front velocity is of the order of $1$~m/sec and is sensitive to field
gradients and sweep rates. We also measured the propagation velocity of a
heat pulse and found that it is much slower than the avalanche velocity. We
therefore conclude that in Fe$_{8}$, the avalanche front propagates without
thermal assistance.
\end{abstract}

\maketitle

Single molecular magnets (SMM) are an excellent model system for the study of
macroscopic quantum phenomena and their interplay with the environment. In
recent years, the focus of these studies shifted from single molecule to
collective effects. While there are two famous SMM that show quantum
behavior, namely, Fe$_{8}$ and Mn$_{12}$, most of the work on collective
effects has been focused on Mn$_{12}$. Indeed, in Mn$_{12}$ intriguing effects
were found, such as deflagration \cite{suzuki2005,Subedi2013}, quantum
assisted deflagration \cite{Hernandez2005}, and detonation \cite{Detonation}%
. In all these cases, a spin reversal front propagates through the sample as
an avalanche. Although showing some signs of quantum behavior \cite%
{Hernandez2005}, these processes are based on over-the-barrier magnetization
reversal. Here, we focus on the spin avalanche phenomena in Fe$_{8}$, where
pure quantum effects exist at dilution refrigerator (DR) temperatures. We
measure the avalanche velocity $V_{a}$ for various sweep rates and applied
field gradients. We also determine the thermal diffusivity. We find that $%
V_{a}$ is much faster than the velocity at which heat or matching field
propagates through the sample. Moreover, $V_{a}$ is affected by field
gradients. Therefore, the avalanche in Fe$_{8}$ is a quantum effect
sometimes called cold deflagration \cite{Garanin2009}. Fe$_{8}$ provides the
first experimental manifestation of such cold deflagration.

The Fe$_{8}$ SMM has spin $S=10$ ground state, as does Mn$_{12}$. The magnetic
anisotropy correspoding to an energy barrier between the spin projection
quantum number $m=\pm 10$ and $m=0$ is $27.5$~K \cite%
{WorensdorferScience99,Mukhin2001,Barra1996,Caciuffo1998,Park2002}; in Mn$_12$
this anisotropy is ~$70$~K \cite{Caneschi,Sessoli}. Fe$_{8}$ molecules show
temperature-independent hysteresis loops at $T<400$~mK, with magnetization
jumps at matching fields that are multiples of $0.225$~T\cite%
{Wernsdorfer2000,Wernsdorfer1999}. However, when tunneling is taking place
from state $m$ to $m^{\prime }$, where $\left\vert m^{\prime }\right\vert
\neq 10$, the excited state can decay to the ground state $\left\vert
m^{\prime }\right\vert =10$, releasing energy in the process. In a
macroscopic sample, this energy release can increase the temperature and
support a deflagration process by assisting the spin flips. Spontaneous
deflagration in Mn$_{12}$ takes place at various and not necessarily
matching fields higher than $1$~T. The deflagration velocity starts from $1$%
~m/sec and increases with an increasing (static) field up to $15$~m/sec \cite%
{suzuki2005}.

\begin{figure}[tbph]
\begin{center}
\includegraphics[clip,
width=\columnwidth]{{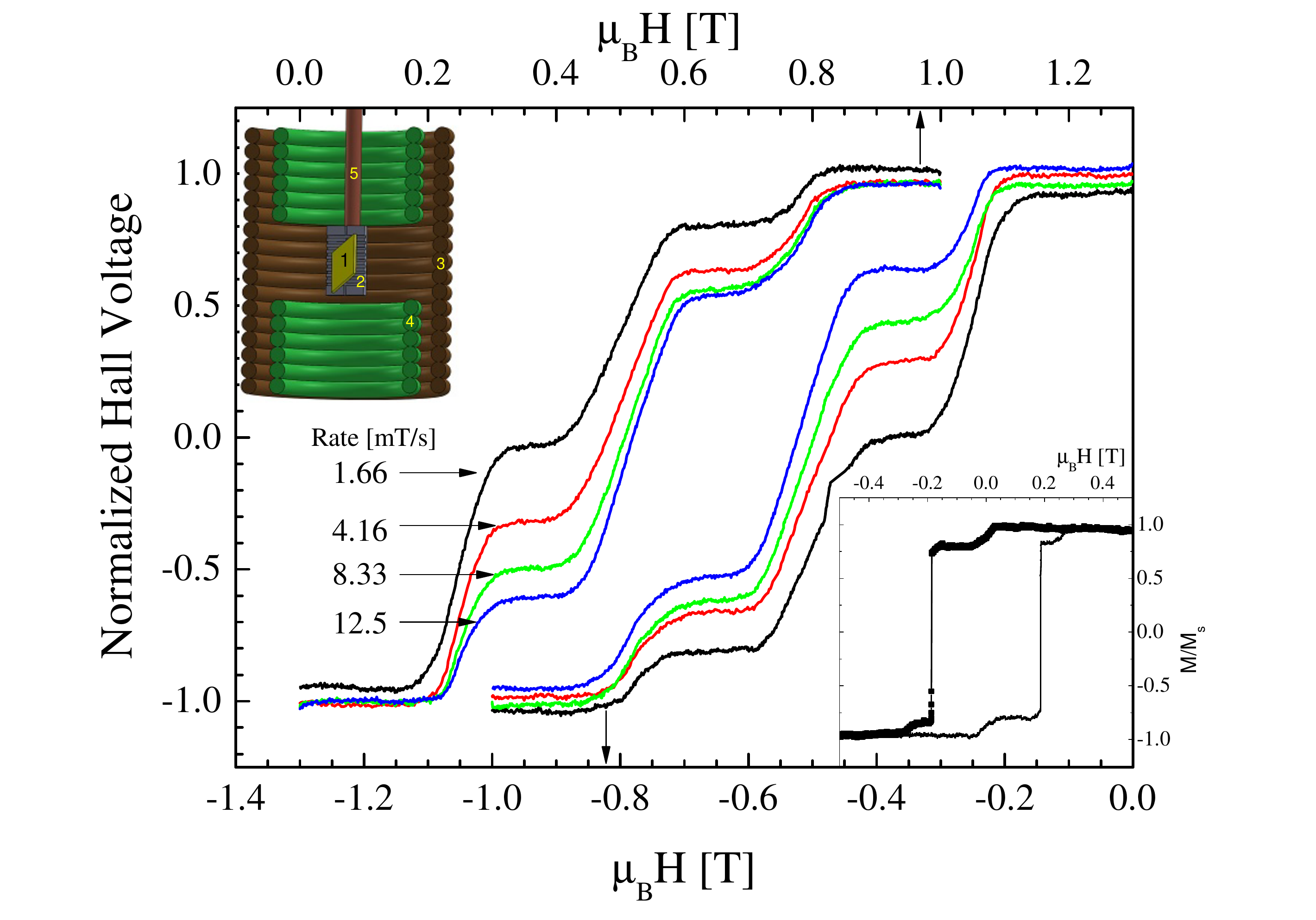}} %HERE Change to Fig1%
\end{center}
\caption{Fe$_{8}$ Hysteresis loops for a sample that does not show
avalanches at different magnetic field sweep rates. The magnetization is
measured via one Hall sensor of the array. The fields for the positive sweep
rates are given by the bottom x-axis, and for the negative sweep rates by
the top x-axis. The upper inset shows the experimental setup including: 1)
sample, 2) Hall sensor array, 3) main coils, 4) gradient coils, and 5) cold
finger leading to the Dilution Refrigerator mixing chamber. The lower inset
shows the hysteresis loop for a sample that does experience avalanches. Only
two magnetization steps are observed in this case.}
\label{Fig1}
\end{figure}

Our avalanche velocity measurements are based on local and time-resolved
magnetization detection using a Hall sensor array. The array is placed at
the center of a magnet and gradient coils. A schematic view of the array and
coils is shown in the inset of Fig.~\ref{Fig1}. The array is made of Hall
bars of dimensions 100$\times $100 $\mu $m$^{2}$ with 100 $\mu $m intervals;
the active layer in these sensors is a two-dimensional electron gas formed
at the interface of GaAs/AlGaAs heterostructures. The surface of the Hall
sensors is parallel to the applied field. Consequently, the effect of the
applied field on the sensor is minimal and determined only by the ability to
align the array surface and field. The sample and sensors are cooled to $100$%
~mK using a DR. More details on the Hall measurements can be found in the
supplemental material.

A magnetic field gradient could also be produced by two superconducting
coils wound in the opposite sense. They are placed at the center of the main
coil and produce $0.14$ mT/mm per ampere. Since there is no option of
adjusting the sample position after it has been cooled it is reasonable to
assume that the sample is not exactly in the center of the main magnet. In
addition, the sample has corners and edges. Therefore, a field gradient is
expected even when the gradient coils are turned off.

In the experiments, the molecules are polarized by applying a magnetic field
of $\pm 1$~T in the $\mathbf{\hat{z}}$ direction. Afterwards, the magnetic
field is swept to $\mp 1$~T. The sweep is done at different sweep rates and
under various applied magnetic field gradients. During the sweep, the
amplified Hall voltage from all sensors and the external field are recorded.
From the raw field-dependent voltage of each sensor, a straight line is
subtracted. This line is due to the response of the Hall sensor to the
external field. The line parameters are determined from very high and very
low fields where no features in the raw data are observed.

In our experiments, we found that Fe$_{8}$ samples can be divided into two
categories: those that do not show avalanches, which have multiple magnetization
steps regardless of the sweep rate, and those that show avalanches where the
number of magnetization steps depends on the sweep rate. In Fig.~\ref{Fig1},
we present the normalized Hall voltage as detected by one of the Hall
sensors from a sample of the first category. The normalization is by the
voltage at a field of 1T where the molecules are fully polarized. Thus, the
normalized voltage provides $M/M_{0}$, where M is the magnetization and $M_{0}$ is the saturation
magnetization. The bottom abscissa is for a sweep where the field decreases
from $1$~T. The top abscissa is for a sweep where the field increases from $%
-1$~T. The magnetization shows typical steps at intervals of 0.225~T. No
step is observed near zero field. In addition, the hysteresis loop's
coercivity increases as the sweep rate increases. These results are in
agreement with previous measurements on Fe$_{8}$ \cite{Wernsdorfer1999}.
They are presented here to demonstrate that the Hall sensors are working
properly, that their signals indeed represent the Fe$_{8}$ magnetization,
and that in some samples all magnetization steps are observed.
\begin{figure}[tbph]
\begin{center}
\includegraphics[clip,
width=\columnwidth]{{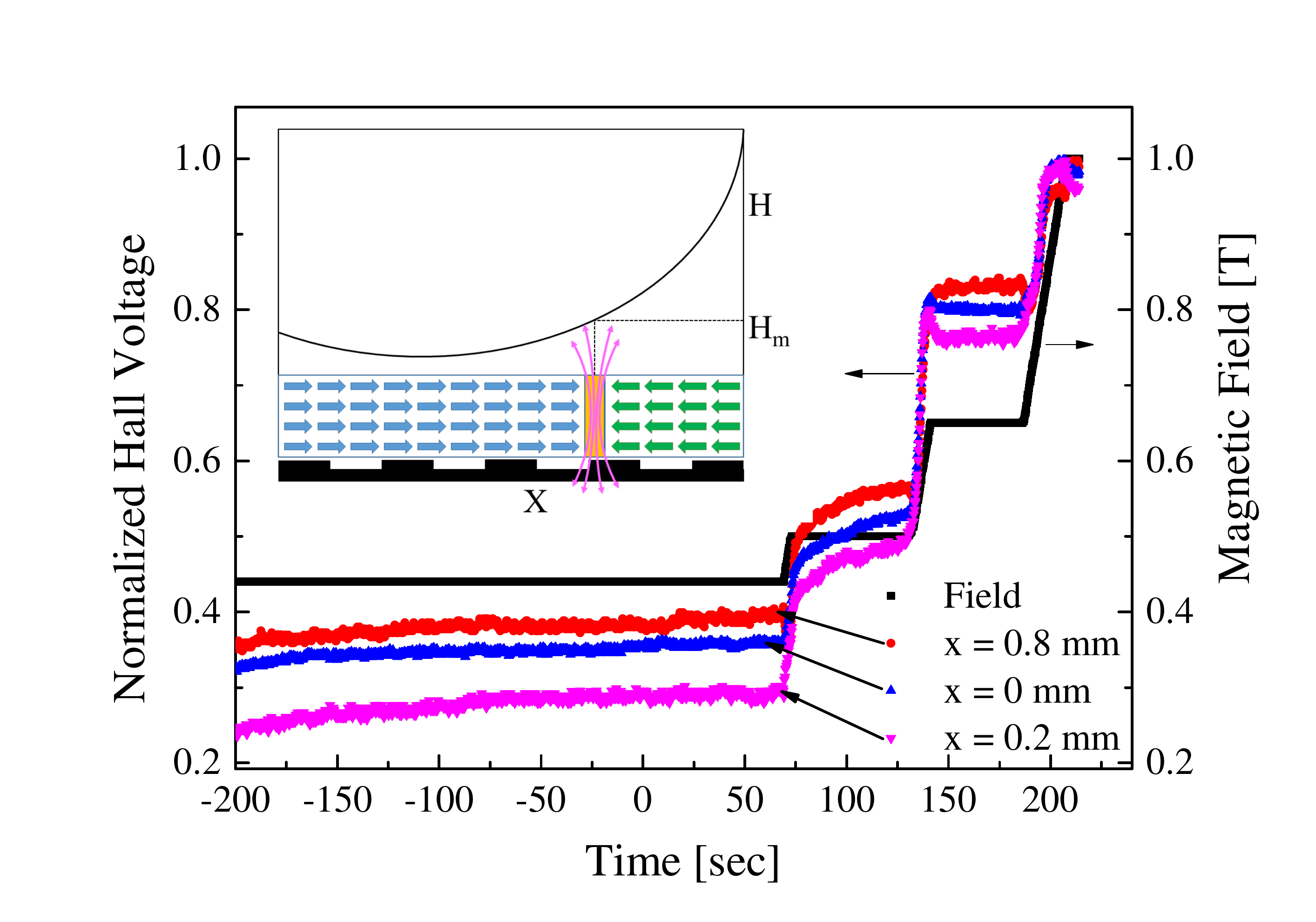}}
\end{center}
\caption{Magnetization as a function of time for a sample of the first type
with no avalanche. The magnetization is measured via three different Hall
sensors. The field is swept discontinuously. The solid (black) line shows
the field value as a function of time on the right y-axis. The
magnetization, presented on the left y-axis, changes only when the field
changes. The inset demonstrates a tunneling front evolution in a case where
the matching field $H_{m}$ moves across the sample during a sweep. $H$ is an
instantaneous field intensity. It changes with time and varies in space. The
tunneling region with mixed up and down spin has zero magnetization. The
expelled magnetic induction $\mathbf{B}$ is detected by the Hall sensors.}
\label{Fig2}
\end{figure}

The hysteresis loop of a sample from the second category is plotted in the
bottom inset of Fig.~\ref{Fig1}. In this case, there is a small
magnetization jump at zero applied field, followed by a nearly full
magnetization reversal at a field of $0.2$~T in the form of an avalanche. In
all samples tested in this and other experiments in our group\cite%
{LeviantPhotons}, avalanches occurred only at the first matching field. We
could not tell in advance whether a sample was of the first or second
category. We always worked with samples of approximately the same dimensions
($3\times 3\times 1$ mm$^{3}$). This is in contrast to Mn$_{12}$, where
avalanches are associated with large samples \cite{Garanin}.

Avalanche velocity measurements in Fe$_{8}$ should be done with extra care.
In an avalanche there is, of course, a propagating front where spins flip.
But since our measurement in Fe$_{8}$ are done by sweeping the field through
resonance, there is a similar front even without avalanche. This is
demonstrated in the inset of Fig.~\ref{Fig2}. In this inset, a sample placed
off the symmetry point of a symmetric field profile is shown. Thus, the
sample experiences a field gradient. Due to this gradient, tunneling of
molecules will start first at a particular point in the sample where the
local field is at matching value. The spin reversal front will then
propagate from that point to the rest of the sample as the external field is
swept. In this case, pausing the field sweep will stop the magnetization
evolution. This is demonstrated in Fig.~\ref{Fig2} for an avalanche free
sample. The left ordinate is the normalized Hall voltage (solid symbols)
from three different sensors on the array. Each symbol represents a
different sensor. The right ordinate is the applied magnetic field (line).
The voltage and field are plotted as a function of time. We focus on fields
before, near, and after the third transition in Fig.~\ref{Fig1}. For the
most part, the magnetization changes only when the field changes, even in
the middle of a magnetization jump. This means that the sample is subjected
to some field gradients and a tunneling front propagates through the sample
even without an avalanche. It is possible to estimate the matching field front
velocity of $V_{m}\sim 1.5\times 10^{-4}$~m/sec from a typical transition width
($0.1$~T), a typical sweep rate ($5$~mT/sec) and the sample length ($3$~mm).

\begin{figure}[tbph]
\begin{center}
\includegraphics[clip,
width=\columnwidth]{{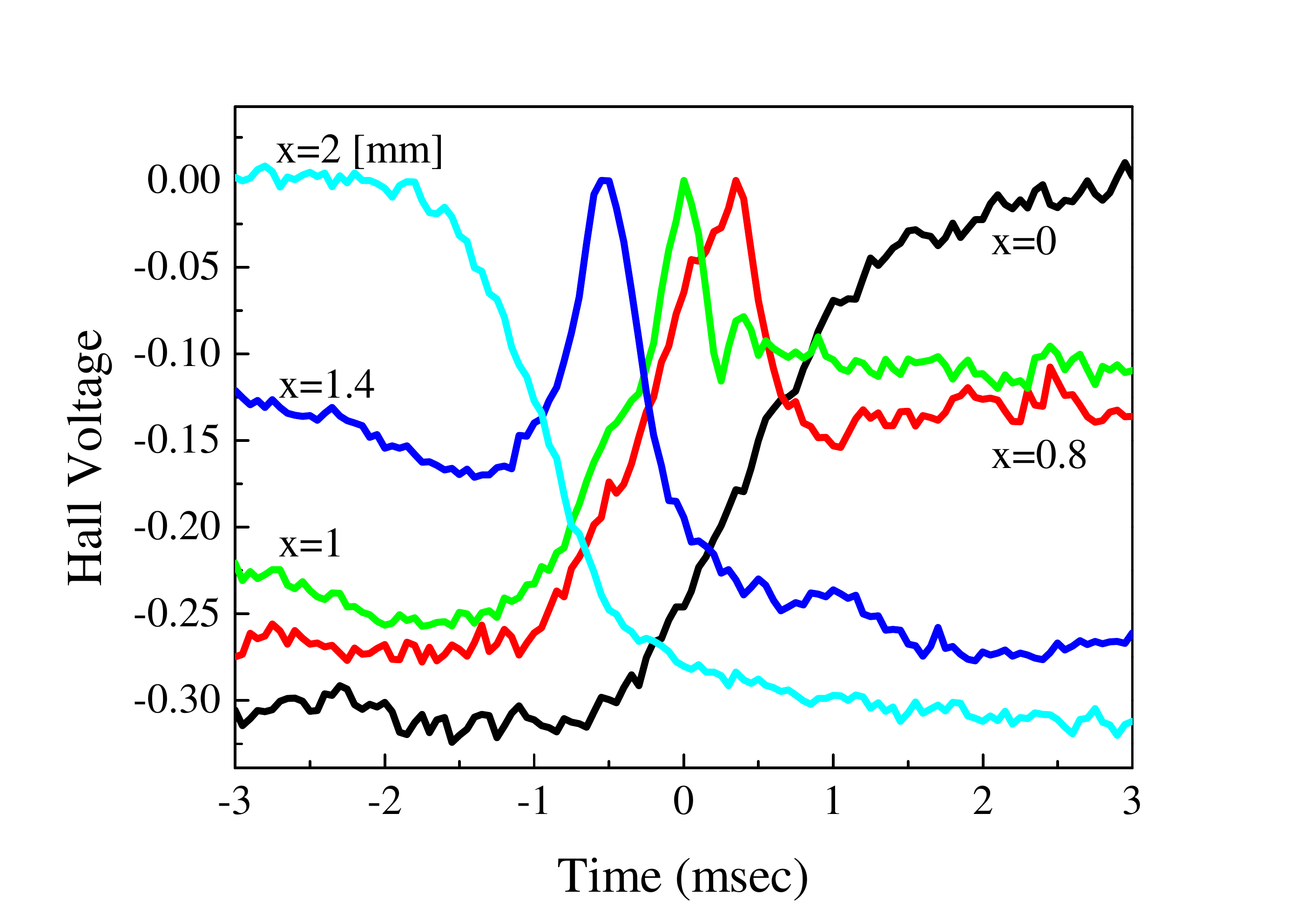}}
\end{center}
\caption{Hall voltage as a function of time for each of the sensors on the
array for a sample that has avalanches (as in the inset of Fig.~\protect\ref%
{Fig2}). The voltage from each sensor shows a peak or a cusp at different
times. The evolution of the peaks and cusps provides the avalanche
propagation velocity.}
\label{Fig3}
\end{figure}

In Fig.~\ref{Fig3}, we zoom in on the magnetization jump of samples from the
second category at a $0.2$~T field. In this figure, we show the
time-resolved Hall voltage from five different sensors along the array. The
three middle sensors show a peak in the Hall voltage, which is experienced
by each sensor at different times. The two outer sensors experience a
smoother variation of the Hall voltage, in the form of cusps, also at
different times. This type of behavior is a clear indication of a
magnetization reversal avalanche propagating from one side of the sample to
the other. The peaks and cusps are due to a zero magnetization front, where
the magnetization $\mathbf{M}$ changes sign due to tunneling. At the same
front, the magnetic induction $\mathbf{B}$ from the sample is forced to
point outward and toward the sensors, to maintain zero divergence \cite%
{Friedman2010}. This is demonstrated in the inset of Fig.~\ref{Fig2}. By
following the time evolution of the peaks and cusps, we can determine the
front velocity. Since the sensors are spaced by parts of a millimeter and
the peaks are spaced by parts of a millisecond, the avalanche velocity $%
V_{a} $ is of the order of $1$~m/sec, which is much higher than $V_{m}$.

\begin{figure}[tbph]
\begin{center}
\includegraphics[clip,
width=0.7\columnwidth]{{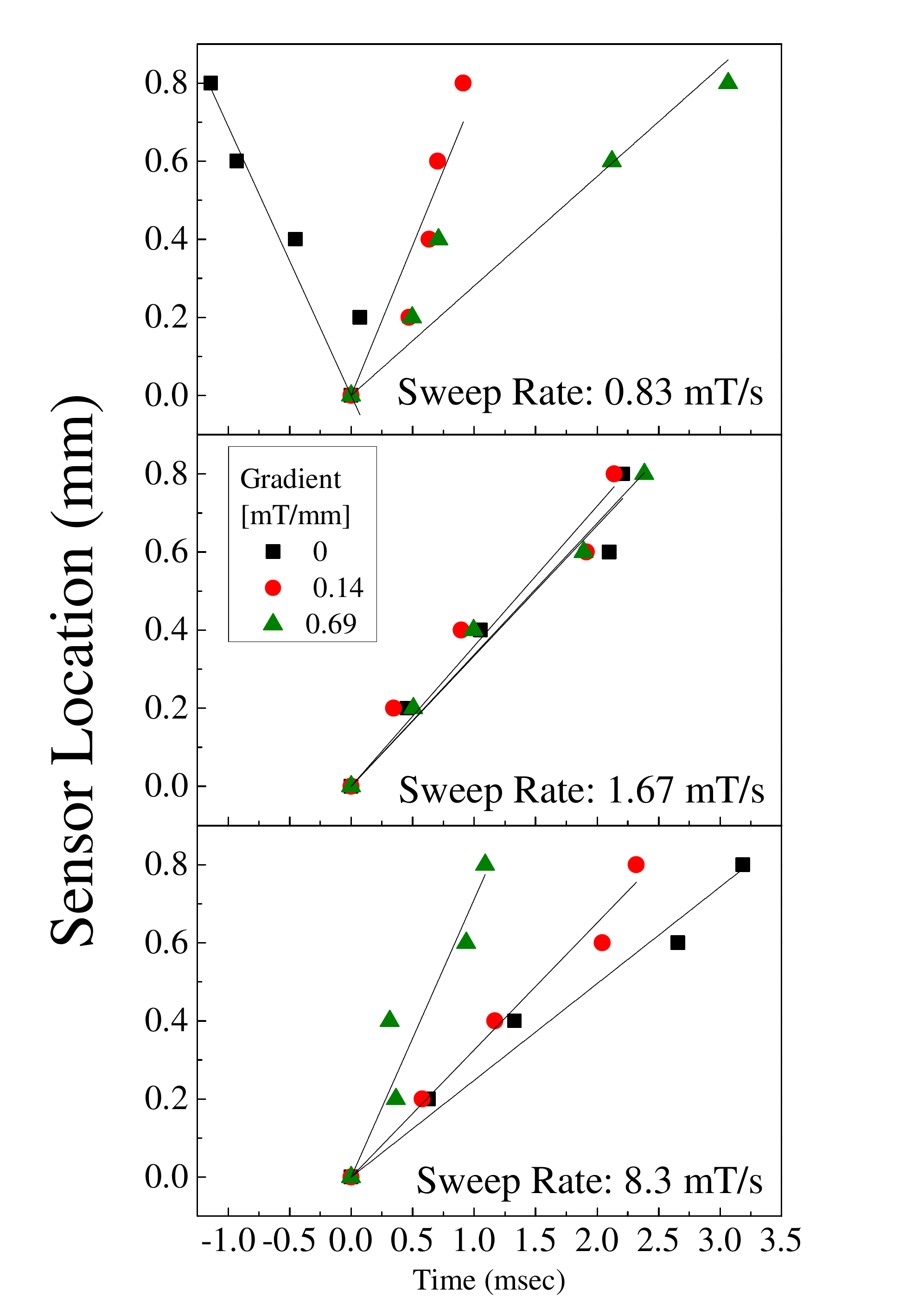}}
\end{center}
\caption{Sensor position as a function of time at which a peak or cusp in
the Hall voltage appears for three different sweep rates and three different
magnetic field gradients. The slope of each line gives the avalanche
velocity. }
\label{Fig4}
\end{figure}

We found that the avalanche propagation direction can be affected by
applying field gradients as long as the sweep rate is low. This is
demonstrated in Fig.~\ref{Fig4}. In this figure, we show for each detector
location the time at which it experiences a peak or a cusp. The slope of
each line is the avalanche velocity. For the lowest sweep rate of
0.83~mT/sec with no gradient, the velocity is negative. It becomes positive
as the gradient is switched on to 0.14 mT/mm, but becomes slower as the
gradient increases to 0.69~mT/mm. The effect of the gradient is opposite and
weaker for our highest sweep rate of 8.3~mT/sec. In this case, all
velocities are positive and increase as the gradient increases. Only at the
intermediate sweep rate of $1.67$~mT/sec does the gradient have no effect on
the velocity. Although we find it challenging to explain the gradient
dependence of the avalanche velocity, we do learn from this experiment that
the safest sweep rate from which one can estimate the avalanche velocity is
around $2$~mT/sec. In this case, the external gradient does not affect the
velocity.

The ratio between sweep rates and gradient (when it is on) is a quantity
with units of velocity of the order of tens of millimeters per second. This
is much lower than $V_{a}$. Therefore, the gradient experiment is another
indication, but with an avalanching sample, that the propagation of the
external magnetic field does not determine the avalanche velocity, and that $%
V_{a}$ is an internal quantity of the molecules. In addition, our ability to
affect $V_{a}$ with the gradient field rules out the possibility that the
avalanche is due to over-the-barrier spin flips.

\begin{figure}[tbph]
\begin{center}
\includegraphics[clip,
width=\columnwidth]{{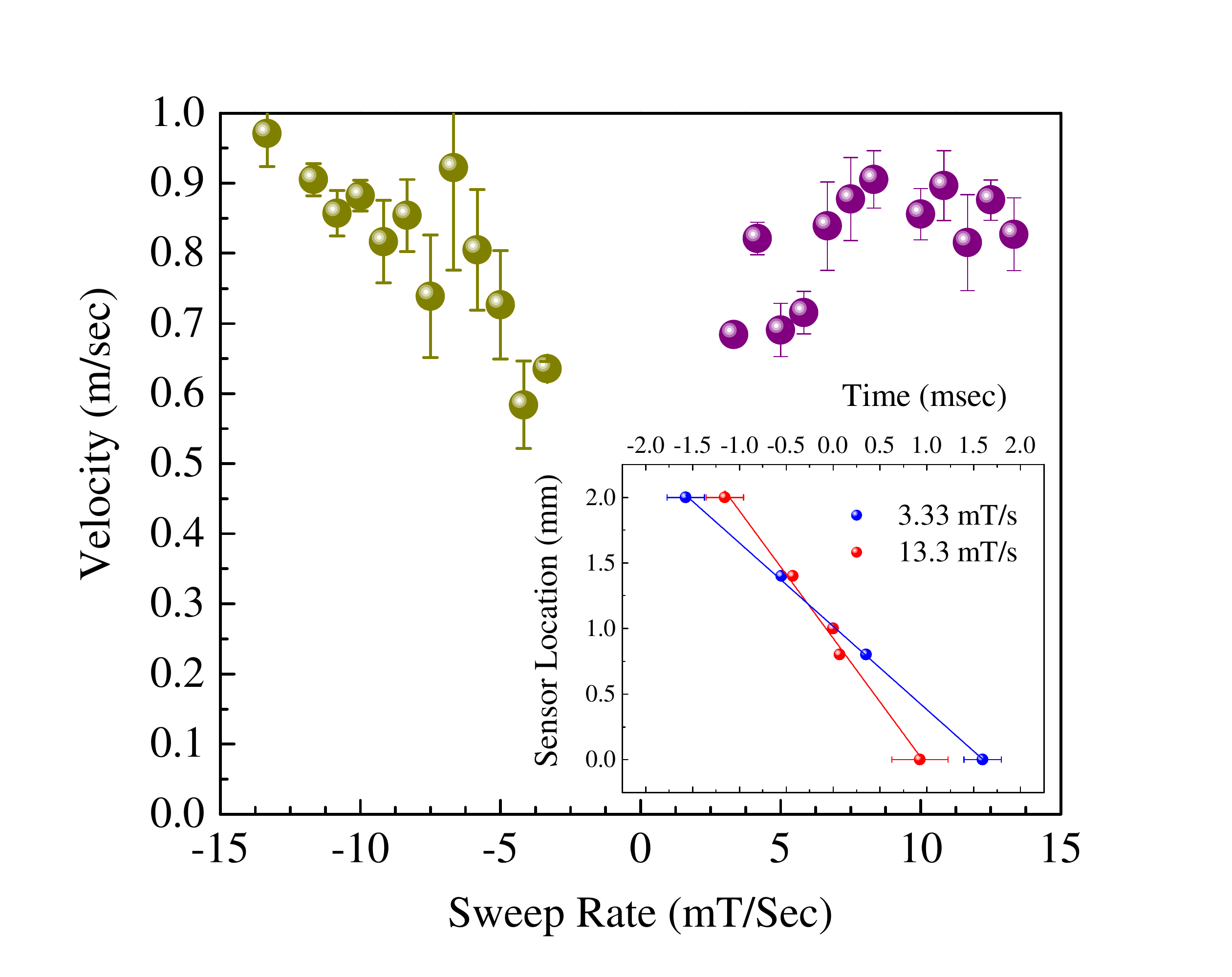}}
\end{center}
\caption{Avalanche velocity as a function of magnetic field sweep rate at
zero gradient. The field is swept from positive to negative and vice versa.
For sweep rates slower than 3 mT/Sec, no avalanche was observed in this
sample. The inset shows raw data of peak position vs. time for two different
sweep rates.}
\label{Fig5}
\end{figure}

Finally, in Fig.~\ref{Fig5} we depict the avalanche velocities $V_{a}$ as a
function of sweep rate with zero applied gradient. The field was swept from
positive to negative and vice versa. The sample used in this experiment was
of the second category and produced avalanches only for sweep rates higher
than 3~mT/sec. Slower sweep rates generated the usual magnetization jumps,
as shown in Fig.~\ref{Fig1}. Although there is some difference between the
velocity for different sweep directions, it is clear that the velocity tends
to increase with increasing sweep rate, and perhaps saturate. In light of
the gradient experiment, the most representative avalanche velocity of Fe$%
_{8}$ is $V_{a}=0.6$~m/sec.

To clarify the role of heat propagation in the avalanche process of Fe$_{8}$%
, we also measured the thermal diffusivity $\kappa $ between $300$~mK and $1$%
~K. This is done by applying a heat pulse on one side of the sample for a
duration of $\tau =1$ msec, and measuring the time-dependent temperature on
the hot side ($T_{hs}$) and on the cold side ($T_{cs}$) of a sample of
length $l\simeq 1$~mm. More experimental details are provided in the supplemental material.
The results are shown in Fig. \ref{Fig6}. The thermal
diffusivity is defined via the heat equation $\frac{\partial T}{\partial t}%
(x,t)-\kappa \frac{\partial ^{2}T(x,t)}{\partial x^{2}}=0$ where $T(x,t)$ is
the location and time dependent temperature along the sample. For a long rod
$\sqrt{\tau \kappa }\ll l$, one has that
\[
\Delta T_{cs}(t)=c\int_{0}^{t}\frac{x\exp
\left( -\frac{x^{2}}{4k(t-s)}\right)}{(4\pi \kappa )^{1/2}(t-s)^{3/2}} \Delta T_{hs}(s)ds.
\]%
We fit this expression to our $T_{cs}(t)$ data with $c$ and $\kappa $ as fit
parameters. $c$ accounts for the coupling of the two thermometers to the
sample. The fit is shown by the solid line in Fig. \ref{Fig6}. Although the
fit is not perfect, it does capture the data quite well. The $\kappa $
obtained with this method at a few different temperatures is depicted in the
inset of Fig.\ref{Fig6}. $\kappa $ and $\tau $ obey the long rod condition.
It is much smaller than $\kappa $ of Mn$_{12}$, which is estimated to be $%
\kappa =10^{-5}$ to $10^{-4}$ m$^{2}$/sec \cite{suzuki2005}. Now, we can
generate a heat velocity $V_{h}=\kappa l/A$ where $A$ is the sample cross
section. At $T=300$~mK we find that $V_{h}=3\times 10^{-3}$ m/sec. This is
roughly $l$ divided by the time between the peak of $T_{hs}(t)$ and that of $%
T_{cs}(t)$.

\begin{figure}[tbph]
\begin{center}
\includegraphics[clip,
width=\columnwidth]{{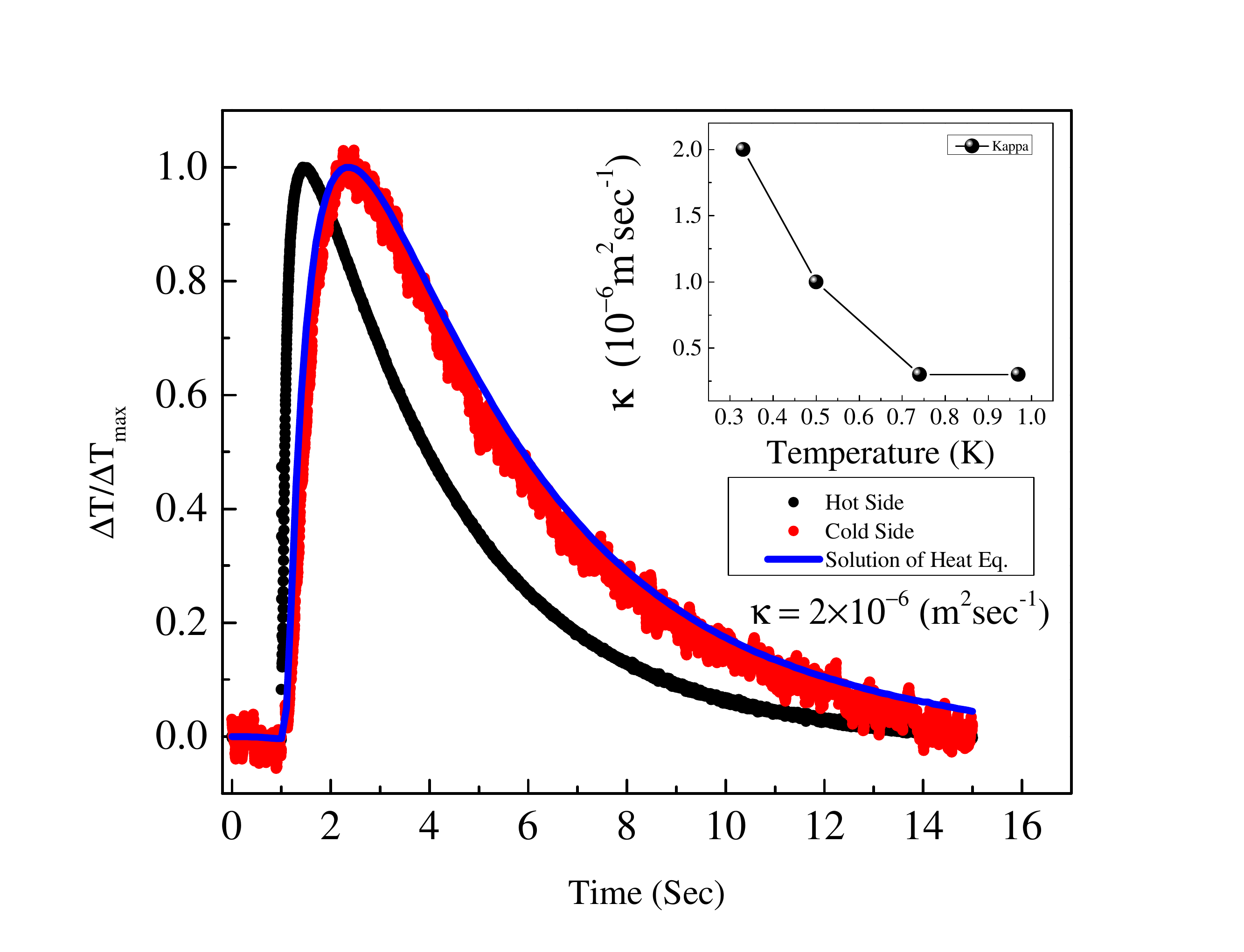}}
\end{center}
\caption{Normalized relative temperature as a function of time at two sides
of the sample. Solid line is solution of heat equation for $\protect\kappa%
=2\times10^{-6}$. The inset shows thermal diffusivity $\protect\kappa $ at
different temperatures}
\label{Fig6}
\end{figure}

Our experiments show that $V_{a}\gg V_{h}>V_{m}$. This means that the spin
reversal front outruns the matching field as it crosses the sample. More
important, the avalanche outruns the heat generated in its wake. Every new
molecular spin that tunnels does so at the DR temperature. Although heat is
produced in the process, this heat does not propel the tunneling front
forward. Moreover, the avalanche starts only at the first matching field and
it's velocity is affected by a field gradient. Therefore, the avalanche
properties are sensitive to the resonance conditions. All these observations
render the avalanche in Fe$_{8}$ a quantum mechanical phenomena. The open
question is then what sets its velocity. A natural guess, of tunnel
splitting $\Delta =4\times 10^{3}$~sec$^{-1}$
times unite cell size of $1.6$~nm, namely, $6\times 10^{-6}$~m/sec is too
slow \cite{WorensdorferScience99}. Therefore, to address this question, more
profound considerations have to be taken into account.

This study was partially supported by the Russell Berrie Nanotechnology
Institute, Technion, Israel Institute of Technology.

\section{Supplemental material}

The Hall sensor array resides in the center of a printed circuit board
(PCB). There is a hole in the PCB and the Hall sensor is glued directly on a
copper plate cold finger, which extends from the DR mixing chamber. Gold
wire bonding connects the sensors and the leads on the PCB. All wires are
thermally connected to the MC. Typical sample dimensions are 3 $\times $ 2 $%
\times $ 1 mm$^{3}$. The samples have clear facets and are oriented with the
easy axis parallel to the applied field. % \cite{suzuki2005}.
They are covered by a thin layer of super glue and placed directly on the
surface of the Hall sensor with Apizon-N grease, which is used to protect
the sample from disintegration and hold it in place. The array backbone has
a resistance of $3-4$ k$\Omega $ at our working temperatures, and is excited
with a 10$~\mu $A DC current. No effect of the sensors' excitation on the
DR-mixing chamber temperature was detected. The Hall voltage from each
sensor is filtered with a 30~Hz low-pass filter for hysteresis measurements
and a 200~Hz high-pass filter for the avalanche measurements. The voltage is
amplified $500$ times by a differential amplifier. It is digitized with an
NI USB 6251 A/D card at a rate of $50$ Hz and $20$ KHz for the hysteresis
and avalanche measurements respectively.

\begin{figure}[tbph]
\begin{center}
\includegraphics[clip,width=0.7\columnwidth]{{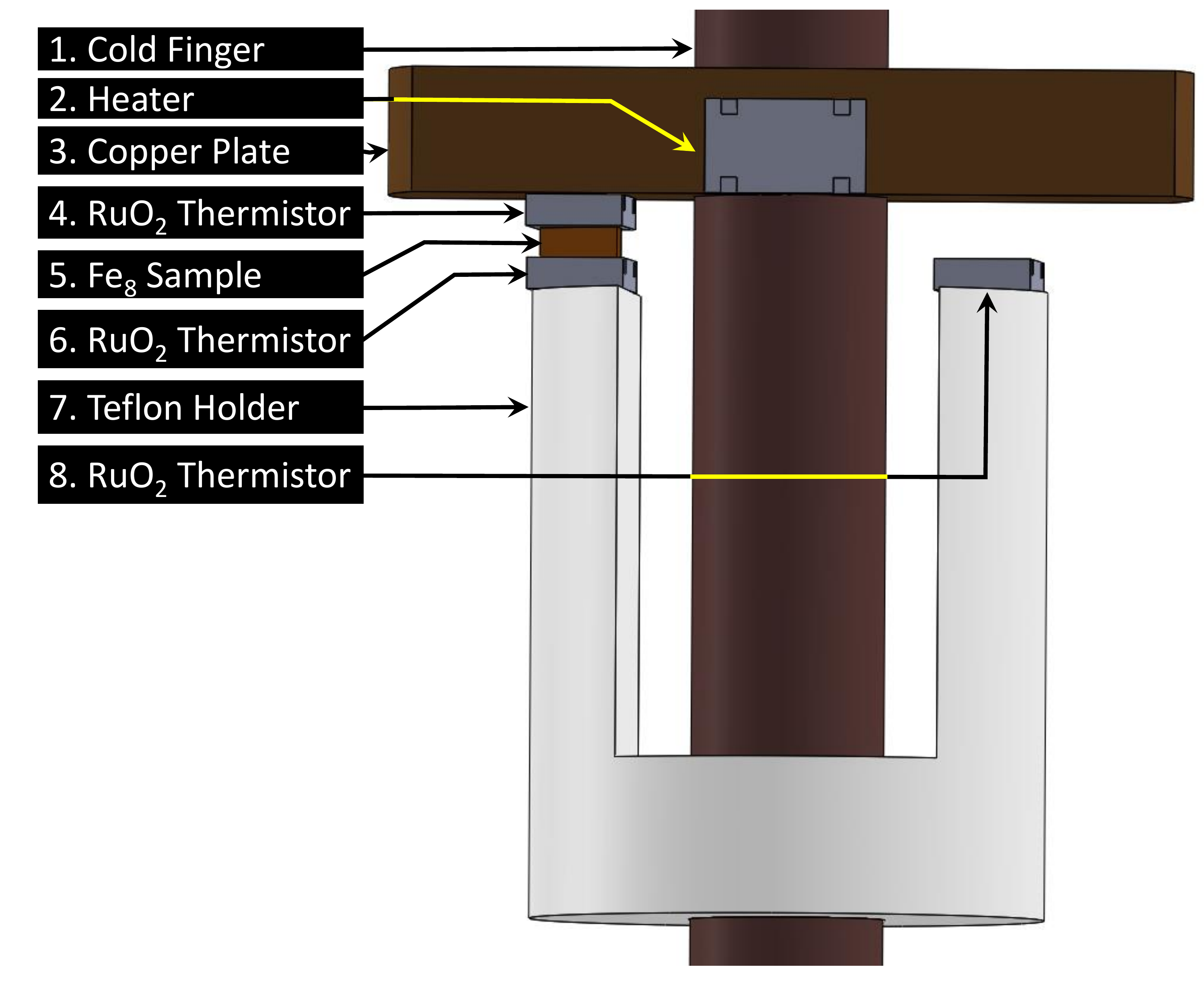}}
\end{center}
\caption{Thermal diffusivity experimental setup. Heat pulse is provided by
heater $2$. Thermistor $4$ measures $T_{hs}$ and Thermistor $6$ measures $%
T_{cs}$. Thermistor $8$ is used to determine heat leaks via the measurement
wires.}
\label{Fig7}
\end{figure}

The thermal diffusivity measurements are performed using two thermometers
mounted on opposite sides of the sample and a heater on the hot side of the
sample,whose configuration is shown in Fig.\ref{Fig7}. The hot side is attached
to the cold finger and is hot only after the heat pulse. The thermometers
are RuO$_{2}$ films. The heater is a 2.2K$\Omega $ resistor. The hot side
thermometer is between the heater and the sample. The cold side thermometer
is between the sample and a teflon plate. It has a weak thermal link to the
cold plate via the measurement wires only. A heat pulse is generated by
applying $8$ V to the 2.2K$\Omega $ resistor using a function generator,
which also gives the trigger for the RuO$_{2}$ voltage measurement. The
system has been tested by repeating the measurement without the sample to
ensure that the recorded heat on the cold side flows through the sample and not
through the wires.

\end{document}